\documentclass[12pt]{article}

\newcommand{\mybar}[1]{\overline{#1}}

\newcommand{\eventra}[1]{\ensuremath{\langle #1 \rangle}}

\def\smallromani{\renewcommand{\theenumi}{\roman{enumi}}
\renewcommand{\labelenumi}{(\theenumi)}}

\newcommand{\Proof}{\NI
                    {\bf Proof.}\ }

\usepackage{amssymb, amsthm}
\usepackage{stmaryrd}
\usepackage{sgame}
\usepackage{url}
\usepackage{amsmath}
\usepackage{handbookbib}

\usepackage{color}
\usepackage{graphics}

\newcommand{\oldbfe}[1]{\begin{bfseries}\emph{#1}\end{bfseries}}

\newcommand{\ES}{\mbox{$\emptyset$}}

\newcommand{\myra}{\mbox{$\:\rightarrow\:$}}

\newcommand{\lra}{\mbox{$\:\leftrightarrow\:$}}

\newcommand{\A}{\mbox{$\ \wedge\ $}}

\newcommand{\sse}{\mbox{$\:\subseteq\:$}}

\newcommand{\fa}{\mbox{$\forall$}}
\newcommand{\te}{\mbox{$\exists$}}

\newcommand{\LL}{\mbox{$\ldots$}}

\newcommand{\newMS}[1]{\mbox{$[\![{#1}]\!]$}}

\newcommand{\C}[1]{\mbox{$\{{#1}\}$}}           

\newcommand{\NI}{\noindent}
\newcommand{\HB}{\hfill{$\Box$}}
\newcommand{\VV}{\vspace{5 mm}}

\newcommand{\II}{\vspace{2 mm}}

\newcommand{\restr}[1]{\! \mid \! {#1}}

\newcommand{\szkew}[1]{\relax \setbox0=\hbox{\kern -24pt $\displaystyle#1$\kern 0pt }%
\box0}
{\catcode`\@=11 \global\let\ifjusthvtest@=\iffalse}

\newcounter{oldmycaption}

\usepackage{amsfonts}
\usepackage{latexsym}
\usepackage{alltt}

\newtheorem{theorem}{Theorem}
\newtheorem{defined}{Definition}
\newenvironment{definition}{\begin{defined} \rm}{\end{defined}}
\newtheorem{exa}{Example}

\newtheorem{lemma}{Lemma}
\newtheorem{corollary}{Corollary}

\newtheorem{note}{Note}

\usepackage{proof}

\usepackage{rotating}

\newcommand{\opt}{\ensuremath{O}}
\newcommand{\rat}{\ensuremath{rat}}
\newcommand{\interp}[1]{{\ensuremath{{\cal I}(#1)}}}
\renewcommand{\implies}{\ensuremath{\myra}}
\newcommand{\eol}{ \vspace{4pt} \\ }

\title{Common Beliefs and Public Announcements in Strategic Games with Arbitrary Strategy Sets}

\author{Krzysztof R. Apt and Jonathan A. Zvesper\footnote{Centrum for
    Mathematics and Computer Science (CWI), Kruislaan 413, 1098 SJ
    Amsterdam, the Netherlands,
    and University of Amsterdam 
}
}

\begin{document}

\maketitle

\begin{abstract}
  We provide an epistemic analysis of arbitrary strategic games based
  on possibility correspondences. We first establish a generic result
  that links true common beliefs (and, respectively, common knowledge)
  of players' rationality defined by means of `monotonic' properties,
  with the iterated elimination of strategies that do not satisfy these
  properties.  It allows us to deduce the customary results concerned
  with true common beliefs of rationality and iterated elimination of
  strictly dominated strategies as simple corollaries.  This approach
  relies on Tarski's Fixpoint Theorem.  
  
  We also provide an axiomatic presentation of this generic result.
  This allows us to clarify the proof-theoretic principles assumed in
  players' reasoning.
  
  Finally, we provide an alternative characterization of the iterated
  elimination of strategies based on the concept of a public
  announcement. It applies to `global properties'.  Both classes of
  properties include the notions of rationalizability and the iterated
  elimination of strictly dominated strategies.
\end{abstract}

\VV

\NI \textbf{Keywords}: epistemic analysis, possibility
correspondences, fixpoints, monotonicity, rationalizability, public
announcements.
\section{Introduction}
\label{sec:intro}

\subsection{Background}

Epistemic analysis of strategic games (in short, games) aims at
predicting the choices of rational players in the presence of (partial
or common) knowledge or belief about the behaviour of other players.
Most often it focusses on the iterated elimination of never best
responses (a notion termed as rationalizability), the iterated
elimination of strictly dominated strategies (IESDS) and on
justification of the strategies selected in Nash and correlated
equilibria.

Starting with \cite{Aum87a}, \cite{BD87} and \cite{TW88} a large body
of literature arose that investigates the epistemic foundations of
rationalizability by modelling the reasoning employed by players in
choosing their strategies.  Such an analysis, based either on
possibility correspondences and partition spaces,
or Harsanyi type spaces, is limited either
to finite or compact games with continuous payoffs, or to two-player
games, see, e.g., \cite{BB99} or \cite{EP06}.

In turn, in the case of IESDS the epistemic analysis has focussed on
finite games (with an infinite hierarchy of beliefs) and strict
dominance either by pure or by mixed strategies, see, e.g.
\cite{BFK08}.

\subsection{Contributions}

In this paper we provide an epistemic analysis of arbitrary strategic
games based on possibility correspondences. We prove a generic result
that is concerned with monotonic program properties\footnote{The
  concepts of monotonic, global and local properties are introduced in
  Section \ref{sec:setup}.} used by the players to select optimal
strategies.

More specifically, given a belief model for the initial strategic
game, denote by $\textbf{RAT}({\overline{\phi}})$ the property that
each player $i$ uses a property $\phi_i$ to select his strategy (`each
player $i$ is $\phi_i$-rational'). 
We establish in Section \ref{sec:epistemic} the following main result:

\begin{quote}
  Assume that each property $\phi_i$ is monotonic.  The
  set of joints strategies that the players choose in the states in which
  $\textbf{RAT}({\overline{\phi}})$ is a true common belief is included
  in the set of joint strategies that remain after the iterated
  elimination of the strategies that for player $i$ are not
  $\phi_i$-optimal.
\end{quote}

In general, transfinite iterations of the strategy elimination are possible.
For some belief models the inclusion can be reversed.

This generic result covers the usual notion of rationalizability in
finite games and a global version of the iterated elimination of
strictly dominated strategies.  For the customary, local version of
the iterated elimination of strictly dominated strategies we justify
in Section \ref{sec:consequences} the statement
\begin{quote}
  true common belief (or common knowledge) of rationality implies that
  the players will choose only strategies that survive the iterated
  elimination of strictly dominated strategies
\end{quote}
for arbitrary games and transfinite iterations of the
elimination process. Rationality refers here to the concept studied in
\cite{Ber84}.

Strict dominance is a non-monotonic property, so the use of monotonic
properties allowed us to provide epistemic foundations for
non-monotonic properties.  However, weak dominance, another
non-monotonic property, remains beyond the reach of this approach.  A
mathematical reason is that its global version is also non-monotonic
(see \cite{Apt07c}), in contrast to strict dominance, the global
version of which is monotonic.  To provide epistemic foundations of
weak dominance the only currently known approach is that of
\cite{BFK08} based on the lexicographic probability systems.

We also provide, in Section \ref{sec:axio}, an axiomatic presentation
of the above generic result.  This clarifies the logical underpinnings of
the epistemic analysis and shows that the use of transfinite
iterations can be naturally captured by a single inference rule that
involves greatest fixpoints.  Also, it shows that the relevant
monotonic properties can be defined using positive formulae.

Finally, inspired by \cite{vB07}, we provide in Section
\ref{sec:public} an alternative characterization of the strategies
that remain after iterated elimination of strategies that for player
$i$ are not $\phi_i$-optimal, based on the concept of a public
announcement due to \cite{Pla89}.  Here monotonicity is not needed and
we obtain a generalization of van Benthem's results to arbitrary
strategic games and to other properties than rationalizability,
notably a global version of weak dominance.

Apart of the necessity of the use of transfinite iterations when
studying arbitrary strategic games, our analysis shows the relevance
of two concepts of the underlying properties $\phi_i$ used by the
players to select their strategies.  The first one is monotonicity
which allows us to use Tarski's Fixpoint Theorem.  The second is
globality, which intuitively means that each subgame obtained by
iterated elimination of strategies is analyzed \emph{in the context}
of the given initial game.  While the proposed epistemic analysis of arbitrary
games based on possibility correspondences crucially depends on the use of
monotonic properties, the one based on public announcement applies to
global properties.

\subsection{Connections}

The relevance of monotonicity in the context of epistemic analysis of 
finite strategic games has already been pointed out in \cite{vB07}.
The distinction between local and global properties is from \cite{Apt07}
and \cite{Apt07c}.

To show that for some belief models an equality holds between the 
sets of joint strategies chosen in  the states in which
  $\textbf{RAT}({\overline{\phi}})$ a true common belief and 
the set of joint strategies that remain after the iterated elimination
  of the strategies that for player $i$ are not $\phi_i$-optimal requires
use of transfinite ordinals.
This complements the findings of \cite{Lip91} in which
transfinite ordinals are used in a study of limited rationality, and
\cite{Lip94}, where a two-player game is constructed for which the
$\omega_0$ (the first infinite ordinal) and $\omega_{0} + 1$ iterations of
the rationalizability operator of \cite{Ber84} differ.  In turn,
\cite{HS98} show that in general arbitrary ordinals are necessary in
the epistemic analysis of strategic games based on partition
spaces.  Further, as argued in \cite{CLL05}, the notion of IESDS \`{a}
la \cite{MR90}, when used for arbitrary games, also requires
transfinite iterations of the underlying operator.

Some of the results presented here were initially reported in a different presentation, in \cite{Apt07a}.

\section{Preliminaries}
\label{sec:prelim}

This paper connects three concepts, 
operators on a complete lattice, strategic games and possibility correspondences.
In this section we introduce these concepts and recall basic results concerning them.

\subsection{Operators}

Consider a fixed complete lattice $(D, \sse)$ with the largest element $\top$.
In what follows we use ordinals and denote them by $\alpha, \beta, \gamma$.
Given a, possibly transfinite, sequence $(G_{\alpha})_{\alpha < \gamma}$ of
elements of $D$ we denote their join and meet respectively by
$\bigcup_{\alpha < \gamma} G_{\alpha}$
and $\bigcap_{\alpha < \gamma} G_{\alpha}$.

\begin{definition}
Let $T$ be an operator on $(D, \sse)$, i.e., $T: D \myra D$.

\begin{itemize}

\item We call $T$ \oldbfe{monotonic} if for all $G, G'$
\[
\mbox{$G \sse G'$ implies $T(G) \sse T(G')$.}
\]

\item We call $T$ \oldbfe{contracting} if for all $G$
\[
T(G) \sse G.
\]

\item We say that an element $G$ is a \oldbfe{fixpoint} of $T$ if $G = T(G)$
and a \oldbfe{post-fixpoint} of $T$ if $G \sse T(G)$.

\item We define by 
transfinite induction a sequence of elements $T^{\alpha}$ of $D$, where $\alpha$ is an ordinal, as follows:

\begin{itemize}

  \item $T^{0} := \top$,

  \item $T^{\alpha+1} := T(T^{\alpha})$,

  \item for all limit ordinals $\beta$, $T^{\beta} := \bigcap_{\alpha < \beta} T^{\alpha}$.
  \end{itemize}

\item We call the least $\alpha$ such that $T^{\alpha+1} = T^{\alpha}$ the \oldbfe{closure ordinal} of $T$
and denote it by $\alpha_T$.  We call then $T^{\alpha_T}$ the \oldbfe{outcome of} (iterating) $T$ and write it alternatively as $T^{\infty}$.
\HB
\end{itemize}
\end{definition}

So an outcome is a fixpoint reached by a transfinite iteration that
starts with the largest element.  In general, the outcome of an
operator does not need to exist but we have the following classic
result due to \cite{Tar55}.\footnote{We use here its `dual' version in
  which the iterations start at the largest and not at the least
  element of a complete lattice.}
\II

\NI
\textbf{Tarski's Fixpoint Theorem} 
Every monotonic operator $T$ on $(D, \sse)$
has an outcome, i.e., $T^{\infty}$ is well-defined.
Moreover,
\[
T^{\infty} = \nu T = \cup \{G \mid G \sse T(G)\},
\]
where $\nu T$ is the largest fixpoint of $T$.
\vspace{2mm}

In contrast, a contracting operator does not need to have a largest fixpoint.
But we have the following obvious observation.

\begin{note} \label{note:con}
Every contracting operator $T$ on $(D, \sse)$ has an outcome, i.e., 
$T^{\infty}$ is well-defined.
\HB
\end{note}

In Section \ref{sec:consequences} we shall need the following lemma, that modifies the corresponding
lemma from \cite{Apt07c} from finite to arbitrary complete lattices.

\begin{lemma} \label{lem:inc}
Consider two operators $T_1$ and $T_2$ on $(D, \sse)$ such that
\begin{itemize}
\item for all $G$, $T_1(G) \sse T_2(G)$,

\item $T_1$ is monotonic,

\item $T_2$ is contracting.
\end{itemize}
Then $T_1^{\infty} \sse T_2^{\infty}$.
\end{lemma}
\Proof
We first prove by transfinite induction that for all $\alpha$
\begin{equation}
T_1^{\alpha} \sse T_2^{\alpha}.
  \label{equ:inc}
\end{equation}

By the definition of the iterations we only need to consider the induction
step for a successor ordinal.  So suppose the claim holds for some
$\alpha$. Then by the first two assumptions and the induction
hypothesis we have the following string of inclusions and equalities:
\[
T_1^{\alpha + 1} =   T_1(T_1^{\alpha}) \sse T_1(T_2^{\alpha}) \sse T_2(T_2^{\alpha}) = T_2^{\alpha + 1}.
\]

This shows that for all $\alpha$ (\ref{equ:inc}) holds.
By Tarski's Fixpoint Theorem and Note \ref{note:con} the outcomes of
$T_1$ and $T_2$ exist, which implies the claim.
\HB

\subsection{Strategic games}

Given $n$ players ($n > 1$) by a \oldbfe{strategic game} (in short, a
\oldbfe{game}) we mean a sequence
$
(S_1, \LL, S_n, p_1, \LL, p_n),
$
where for all $i \in \{1, \LL, n\}$

\begin{itemize}
\item $S_i$ is the non-empty set of \oldbfe{strategies} (sometimes called \oldbfe{actions})
available to player $i$,

\item $p_i$ is the \oldbfe{payoff function} for the  player $i$, so
$
p_i : S_1 \times \LL \times S_n \myra \cal{R},
$
where $\cal{R}$ is the set of real numbers.
\end{itemize}

We denote the strategies of player $i$ by $s_i$, possibly with some
superscripts.  Given $s \in S_1 \times \LL \times S_n$ we denote the
$i$th element of $s$ by $s_i$, write sometimes $s$ as $(s_i, s_{-i})$,
and use the following standard notation:

\begin{itemize}
\item $s_{-i} := (s_1, \LL, s_{i-1}, s_{i+1}, \LL, s_n)$,

\item $S_{-i} := S_1 \times \LL \times S_{i-1} \times S_{i+1} \times \LL \times S_n$.

\end{itemize}

Given a finite non-empty set $A$ we denote by
$\Delta A$ the set of probability distributions over $A$ and call
any element of $\Delta S_i$ a \oldbfe{mixed strategy} of player $i$.

In the remainder of the paper we assume an initial strategic game
\[
H := (H_1, \LL, H_n, p_1, \LL, p_n).
\]
A \oldbfe{restriction} of $H$ is a sequence $(G_1, \LL, G_n)$ such that
$G_i \sse H_i$  for all $i \in \{1, \LL, n\}$. We identify the restriction 
$(H_1, \LL, H_n)$ with $H$.
We shall focus on the complete lattice
that consists of the set of all restrictions of the game $H$
ordered by the componentwise set inclusion:
\[
\mbox{$(G_1, \LL, G_n) \sse (G'_1, \LL, G'_n)$ iff $G_i \sse G'_i$ for all $i \in \{1, \LL, n\}$.}
\]
So $H$ is the largest element in this lattice and
$\bigcup_{\alpha < \gamma}$
and $\bigcap_{\alpha < \gamma}$ are the customary set-theoretic 
operations on the restrictions.

Consider now a restriction $G := (G_1, \LL, G_n)$ of $H$
and two strategies $s_i, s'_i$ from $H_i$ (so \emph{not necessarily} from $G_i$). 
We say that 
$s_i$ \oldbfe{is strictly dominated} \oldbfe{on} $G$ by $s'_i$
(and write $s'_i \succ_{G} s_i$)
if
\[
\fa s_{-i} \in G_{-i} \: p_{i}(s'_i, s_{-i}) > p_{i}(s_i, s_{-i}),
\]
and that $s_i$ \oldbfe{is weakly dominated} \oldbfe{on} $G$ by $s'_i$
(and write $s'_i \succ^{w}_{G} s_i$)
if
\[
\fa s_{-i} \in G_{-i} \: p_{i}(s'_i, s_{-i}) \geq p_{i}(s_i, s_{-i}) \A \te s_{-i} \in G_{-i} \: p_{i}(s'_i, s_{-i}) > p_{i}(s_i, s_{-i}).
\]

In the case of finite games, once the payoff function is extended in 
the expected way to mixed strategies, the relations $\succ_{G}$ and  $\succ^{w}_{G}$ between
a mixed strategy and a pure strategy are defined in the same way.

A \oldbfe{belief} of player $i$ held in $G := (G_1, \LL, G_n)$ can be 

\begin{itemize}
\item a joint strategy
of the opponents of player $i$ in $G$ (i.e., $s_{-i} \in G_{-i}$), 

\item or, in the case the game is finite, a joint mixed strategy of
  the opponents of player $i$ (i.e., $(m_1, \LL, m_{i-1}, m_{i+1},
  \LL, m_n)$, where $m_j \in \Delta G_j$ for all $j$),

\item or, in the case the game is finite, a \oldbfe{correlated strategy} of the opponents of player $i$
  (i.e., $m \in \Delta G_{-i}$).
\end{itemize}

Each payoff function $p_i$ can be modified to an
\oldbfe{expected payoff} function $p_i : G_i \times {\cal B}_{i} \myra
\cal{R}$, where ${\cal B}_{i}$ is one of the above three sets of beliefs of player $i$.

Further, given a restriction $G' := (G'_1, \LL, G'_n)$ of $H$,
we say that 
the strategy $s_i$ from $H_i$ is a \oldbfe{best response in
$G'$ to some belief $\mu_i$ held in $G$}
if 
\[
\fa s'_i \in G'_i \:
p_i(s_i, \mu_i) \geq p_i(s'_i, \mu_i).
\]

\subsection{Possibility correspondences}
\label{subsec:poss}

In this and the next subsection we essentially follow the exposition of \cite{BB99}.
Fix a non-empty set $\Omega$ of \oldbfe{states}.
By an \oldbfe{event} we mean a subset of $\Omega$.

A \oldbfe{possibility correspondence} is a mapping from $\Omega$ to 
the powerset ${\cal P}(\Omega)$ of $\Omega$.
We consider three properties of a possibility correspondence $P$:

\begin{enumerate}\smallromani
\item for all $\omega$, $P(\omega) \neq \ES$,

\item for all $\omega$ and $\omega'$, $\omega' \in P(\omega)$ implies $P(\omega') = P(\omega)$,

\item for all $\omega$, $\omega \in P(\omega)$.
\end{enumerate}

If the possibility correspondence satisfies properties (i) and (ii),
we call it a \oldbfe{belief correspondence} and if it satisfies
properties (i)--(iii), we call it a \oldbfe{knowledge
  correspondence}.\footnote{Note that the notion of a belief has two
  meanings in the literature on epistemic analysis of strategic games,
  so also in this paper.  From the context it is always clear which
  notion is used.  In the modal logic terminology a belief
  correspondence is a frame for the modal logic KD45 and a knowledge
  correspondence is a frame for the modal logic S5, see, e.g.
  \cite{BRV01}.}  Note that each belief correspondence $P$ yields a
partition $\{P(\omega) \mid \omega \in \Omega\}$ of $\Omega$.

Assume now that each player $i$ has at its disposal a possibility
correspondence $P_i$.  
Fix an event $E$. We define 
\[
\square E := \square^1 E := \{\omega \in \Omega \mid
\fa i \in \{1, \LL, n\} \: P_i(\omega) \sse E\},
\]
by induction on $k \geq 1$
\[
\square^{k+1} E := \square \square^{k} E,
\]
and finally
\[
\square^* E := \bigcap_{k =1}^{\infty} \square^{k} E.
\]

If all $P_i$s are belief correspondences, we usually write $B$ instead
of $\square$ and if all $P_i$s are knowledge correspondences, we
usually write $K$ instead of $\square$.  When $\omega \in B^* E$, we
say that the event $E$ is \oldbfe{common belief in the state $\omega$}
and when $\omega \in K^* E$, we say that the event $E$ is
\oldbfe{common knowledge in the state $\omega$}.

By property (iii) of the possibility correspondences we have $K E \sse E$ 
and $K^* E \sse E$.

An event $F$ is called \oldbfe{evident} if $F \sse \square F$.  That
is, $F$ is evident if for all $\omega \in F$ we have $P_i(\omega) \sse
F$ for all $i \in \{1, \LL, n\}$.  In what follows we shall use the
following alternative characterizations of common belief and common
knowledge based on the evident events:
\begin{equation}
  \begin{array}{l}
\mbox{$\omega \in \square^* E$ iff for some evident event $F$ we have $\omega \in F \sse \square E$,} 
  \end{array}
\label{equ:cb}
\end{equation}
where $\square = B$ or $\square = K$ (see \cite{MS89}, respectively
Proposition 4 on page 180 and Proposition on page 175),
\begin{equation}
  \label{equ:ck}
\mbox{$\omega \in K^* E$ iff for some evident event $F$ we have $\omega \in F \sse E$,}
\end{equation}
(see \cite[ page 1237]{Aum76}).

Finally, in Section \ref{sec:axio} we shall use the following 
alternative characterization of common beliefs and common knowledge.

\begin{note} \label{not:alt}
For all belief correspondences
\[
\square^* E := \bigcup \{ F \subseteq \Omega \mid F \subseteq  \square (E \cap F).
\]
\end{note}
\Proof
We have the following string of equivalences:
\[
\begin{array}{lll}
\mbox{$F$ is evident and $F \sse \square E$} & \textrm{iff} & \mbox{$F \sse \square F$ and $F \sse \square E$} \\
                                             & \textrm{iff} & \mbox{$F \sse \square F \cap \square E$} \\
                                             & \textrm{iff} & \mbox{$F \sse \square (F \cap E)$},
\end{array}
\]
so the claim follows by (\ref{equ:cb}).
\HB

\subsection{Models for games}

We now link these considerations with the strategic games.  Given a
restriction $G := (G_1, \LL, G_n)$ of the initial game $H$, by a
\oldbfe{model} for $G$ we mean a set of states $\Omega$ together with
a sequence of functions $\mybar{s_i}: \Omega \myra G_i$, where $i \in
\{1, \LL, n\}$. We denote it by $(\Omega, \mybar{s_1}, \LL, \mybar{s_n})$.

In what follows, given a function $f$ and a subset $E$ of
its domain, we denote by $f(E)$ the range of $f$ on $E$ and by $f
\restr{E}$ the restriction of $f$ to $E$. 

By the \oldbfe{standard model} ${\cal M}$ for $G$ we mean the model in which

\begin{itemize}
 \item $\Omega := G_1 \times \LL \times G_n$ (which means that for
   $\omega \in \Omega$, $\omega_i$ is well-defined),

 \item $\mybar{s_i}(\omega) := \omega_i$.
\end{itemize}
So the states of the standard model for $G$ are exactly the joint strategies in $G$,
and each $\mybar{s_i}$ is a projection function.
Since the initial game $H$ is given, we know the payoff functions $p_1,
\LL, p_n$. So in the context of $H$ a standard model is just an alternative way of
representing a restriction of $H$.

Given a (not necessarily standard) model ${\cal M} := (\Omega, \mybar{s_1}, \LL, \mybar{s_n})$ for a restriction
$G$ and a vector of events $\overline{E} = (E_1, \LL, E_n)$ in ${\cal
  M}$ we define
\[
G_{\overline{E}} := (\mybar{s_1}(E_1), \LL, \mybar{s_n}(E_n))
\]
and call it the \oldbfe{restriction of $G$ to $\overline{E}$}.
When each $E_i$ equals $E$ we write $G_{E}$ instead of $G_{\overline{E}}$.

Finally, we extend the notion of a model for a restriction $G$ to a
\oldbfe{belief model} for $G$ by assuming that each player $i$ has a
belief correspondence $P_i$ on $\Omega$. If each $P_i$ is a knowledge
correspondence, we refer then to a \oldbfe{knowledge model}.

\section{Local and global properties}
\label{sec:setup}

The assumption that each player is rational is one of the basic
stipulations within the framework of strategic games. However,
rationality can be differently interpreted by different
players.\footnote{This matter is obfuscated by the fact that the
  etymologically related noun `rationalizability' stands by now for
  the concept introduced in \cite{Ber84} and \cite{Pea84} that refers
  to the outcome of iterated elimination of never best responses.} This
may for example mean that a player

\begin{itemize}
\item does not choose a strategy weakly/strictly dominated by another pure/mixed strategy,
  
\item chooses only best replies to the (beliefs about the) strategies
  of the opponents.
\end{itemize}

In this paper we are interested in analyzing situations in which each
player pursues his own notion of rationality, more specifically those
situations in which this information is common knowledge or common
belief.  As a special case we cover then the usually analyzed
situation in which all players use the same notion of rationality.

Given player $i$ in the initial strategic game $H := (H_1, \LL, H_n, p_1,
\LL,p_n)$ we formalize his notion of rationality using a property
$\phi_i(s_i, G)$ that holds between a strategy $s_i \in H_i$ and a
restriction $G$ of $H$.  Intuitively, $\phi_{i}(s_i, G)$
holds if $s_i$ is an `optimal' strategy for player $i$ within the
restriction $G$, assuming that he uses the property
$\phi_i$ to select optimal strategies.

We distinguish though between what we call `local' and `global' optimality.
To assess optimality of a strategy $s_i$ \emph{locally} within the restriction $G$,
it is sufficient for $i$ to compare $s_i$ with \emph{only those strategies} $s'_i$
\emph{that occur in} $G_i$.  On the other hand, to assess the optimality of
$s_i$ \emph{globally}, player $i$ must consider all of his strategies $s'_i$ that occur
in his strategy set $H_i$ in the \emph{initial game} $H$.

Global properties are then those in which a player's strategy is evaluated
with respect to all his strategies in the initial game, whereas local properties
are concerned solely with a comparison of strategies available in the restriction $G$.
We will write $\phi^l$ when we refer to a local property, and $\phi^g$
when we refer to a global property.

Here are some examples which show that the notions of rationality mentioned above
can be formalized in a number of natural ways.  We also give one example in both its
local and global form in order to illustrate the distinction between them:

\begin{itemize}
  
\item $sd^l_{i}(s_i, G)$ that holds iff the strategy $s_i$ of
  player $i$ is not strictly dominated on $G$ by any strategy from $G_i$
  (i.e., $\neg \te s'_i  \in G_i \: s'_i \succ_{G} s_i$),

\item $sd^g_{i}(s_i, G)$ that holds iff the strategy $s_i$ of
  player $i$ is not strictly dominated on $G$ by any strategy from $H_i$
  (i.e., $\neg \te s'_i  \in H_i \: s'_i \succ_{G} s_i$),

\item (assuming $H$ is finite) $msd^l_{i}(s_i, G)$ that holds iff
  the strategy $s_i$ of player $i$ is not strictly dominated on $G$ by
  any of its mixed strategy from $G$, (i.e., $\neg \te m'_i \in \Delta G_i \: m'_i
  \succ_{G} s_i$),

\item $wd^l_{i}(s_i, G)$ that holds iff the strategy $s_i$ of
  player $i$ is not weakly dominated on $G$ by any strategy from $G_i$
  (i.e., $\neg \te s'_i \in G_i \: s'_i \succ^{w}_{G} s_i$),

\item (assuming $H$ is finite) $mwd^l_{i}(s_i, G)$ that holds iff
  the strategy $s_i$ of player $i$ is not weakly dominated on $G$ by
  any mixed strategies over $G_i$
  (i.e., $\neg \te m'_i \in \Delta G'_i \: m'_i \succ^{w}_{G} s_i$),

\item $br^l_{i}(s_i, G)$ that holds iff the strategy $s_i$ of
  player $i$ is a best response among $G_i$ to some belief
  $\mu_i$ held in $G$ (i.e., for some belief $\mu_i$ held in $G$,
  $\fa s'_i \in G_i \: p_i(s_i, \mu_i) \geq p_i(s'_i, \mu_i)$).
\end{itemize}

We say that the property $\phi_{i}(\cdot, \cdot)$ used by player $i$
is \oldbfe{monotonic} if
for all restrictions $G$ and $G'$ of $H$ and $s_i \in H_i$
\[
\mbox{$G \sse G'$ and $\phi(s_i, G)$ implies $\phi(s_i, G')$.}
\]

Each sequence of properties $\overline{\phi} := (\phi_1, \LL, \phi_n)$
determines an operator $T_{\overline{\phi}}$ on the restrictions of
$H$ defined by
\[
T_{\overline{\phi}}(G) := (G'_1, \LL, G'_n),
\]
where $G := (G_1, \LL, G_n)$ and for all $i \in \{1, \LL, n\}$
\[
G'_i := \{ s_i \in G_i \mid \phi_i(s_i, G)\}.
\]

Since $T_{\overline{\phi}}$ is contracting, by Note \ref{note:con} it
has an outcome, i.e., $T_{\overline{\phi}}^{\infty}$ is well-defined.
Moreover, if each $\phi_i$ is monotonic, then $T_{\overline{\phi}}$ is
monotonic and by Tarski's Fixpoint Theorem its largest fixpoint $\nu
T_{\overline{\phi}}$ exists and equals $T_{\overline{\phi}}^{\infty}$.

Intuitively, $T_{\overline{\phi}}(G)$ is the result of removing from
$G$ all strategies that are not $\phi_i$-optimal. So the outcome of
$T_{\overline{\phi}}$ is the result of the iterated elimination of
strategies that for player $i$ are not $\phi_i$-optimal, where $i \in
\{1, \LL, n\}$.

When each property $\phi_i$ equals ${\textit{sd}^{\: l}}$, we write
$T_{{\textit{sd}^{\: l}}}$ instead of $T_{\overline{{\textit{sd}^{\:
        l}}}}$ and similarly with other specific properties.  The
natural examples of such an iterated elimination of strategies that
were discussed in the literature are:\footnote{The reader puzzled by
  the existence of multiple definitions for the apparently uniquely
  defined concepts is encouraged to consult \cite{Apt07}.}

\begin{itemize}
\item iterated elimination of strategies that are strictly dominated by another strategy;
  
  This corresponds to the iterations of the $T_{\textit{sd}^{\: l}}$
  operator in the case of \cite{DS02}) and of the $T_{\textit{sd}^{\:
      g}}$ operator in the case of \cite{CLL05}.

\item iterated elimination of strategies that are weakly dominated by another strategy;

\item (for finite games)
iterated elimination of strategies that are weakly, respectively strictly, dominated by a mixed strategy;

These are the customary situations studied starting with \cite{LR57} that
correspond to the iterations of the $T_{\textit{msd}^{\: l}}$, respectively $T_{\textit{mwd}^{\: l}}$,
operator.

\item iterated elimination of strategies that are never best responses to some belief;

This corresponds to the iterations of the $T_{\textit{br}^{\: g}}$ operator
in the case of \cite{Ber84} and the $T_{\textit{br}^{\: l}}$ operator
in the case of \cite{Pea84}, in each case for an appropriate set of beliefs.

\end{itemize}

Usually only the first $\omega_o$ iterations of the corresponding
operator $T$ are considered, i.e., one studies
$T^{\omega_0}$, that is $\bigcap_{i< \omega_0}
T^{i}$, and not $T^{\infty}$.

In the next section we assume that each player $i$ employs some
property $\phi_i$ to select his strategies, and we analyze the situation
in which this information is tru common belief or common knowledge.  To determine which
strategies are then selected by the players we shall use the
$T_{\overline{\phi}}$ operator.  We shall also explain why in general
transfinite iterations are necessary.

\section{Two theorems}
\label{sec:epistemic}


Fix a belief model 
$(\Omega, \mybar{s_1}, \LL, \mybar{s_n}, P_1, \LL, P_n)$ for the initial game $H$.
Given a property $\phi_i(\cdot, G)$ that player $i$ uses to select
his strategies in the restriction $G$ of $H$, we say that 
player $i$ is $\phi_i$-\oldbfe{rational in the
  state} $\omega$ if $\phi_i(\mybar{s_{i}}(\omega), G_{P_i(\omega)})$ holds.
Note that when player $i$ believes (respectively, knows) that the
state is in $P_i(\omega)$, the restriction $G_{P_i(\omega)}$
represents his belief (respectively, his knowledge) about the players'
strategies.  That is, $G_{P_i(\omega)}$ is the game he believes
(respectively, knows) to be relevant to his choice.
Hence $\phi_i(\mybar{s_{i}}(\omega), G_{P_i(\omega)})$ captures the idea that
if player $i$ uses $\phi_i(\cdot, \cdot)$ to select his optimal
strategy in the game he considers relevant, then in the state $\omega$
he indeed acts `rationally'.

We are interested in the strategies selected by each player in the
states in which it is true and is common belief (or is common knowledge)
that each player $i$ is $\phi_i$-rational.  To this end we
introduce the following event:
\[
\mbox{$\textbf{RAT}({\overline{\phi}}) := \{\omega \in \Omega \mid $ each player $i$ is $\phi_i$-rational in $\omega$\},}
\]
and consider the following two events constructed out of it:
$K^* \textbf{RAT}({\overline{\phi}})$ and 
$\textbf{RAT}({\overline{\phi}}) \cap B^* \textbf{RAT}({\overline{\phi}})$.
We then focus on the corresponding restrictions $G_{K^* \textbf{RAT}({\overline{\phi}})}$ 
and $G_{\textbf{RAT}({\overline{\phi}}) \cap B^* \textbf{RAT}({\overline{\phi}})}$.

So a strategy $s_i$ is an element of the $i$th component of 
$G_{K^* \textbf{RAT}({\overline{\phi}})}$ if
$s_i = \mybar{s_i}(\omega)$ for some
$\omega \in K^* \textbf{RAT}({\overline{\phi}})$. That is,
$s_i$ is a strategy that player $i$ chooses in a state in which 
it is common knowledge that each player $j$ is $\phi_j$-rational, and similarly for 
$G_{\textbf{RAT}({\overline{\phi}}) \cap B^* \textbf{RAT}({\overline{\phi}})}$.

The following result then relates for arbitrary strategic games the
restrictions $G_{\textbf{RAT}({\overline{\phi}}) \cap B^*
  \textbf{RAT}({\overline{\phi}})}$ and $G_{K^*
  \textbf{RAT}({\overline{\phi}})}$ to the outcome of the iteration of
the operator $T_{\overline{\phi}}$.

\begin{theorem} \label{thm:epist1}
\mbox{}\\[-6mm]
\begin{enumerate} \smallromani
\item Suppose that each property $\phi_i$ is monotonic. Then for all belief models for $H$
\[
G_{\textbf{RAT}({\overline{\phi}}) \cap B^* \textbf{RAT}({\overline{\phi}})} \sse T_{\overline{\phi}}^{\infty}.
\]

\item Suppose that each property $\phi_i$ is monotonic. Then for all knowledge models for $H$
\[
G_{K^* \textbf{RAT}({\overline{\phi}})} \sse T_{\overline{\phi}}^{\infty}.
\]

\item For some standard belief model for $H$
\[
G_{\textbf{RAT}({\overline{\phi}}) \cap B^* \textbf{RAT}({\overline{\phi}})} = T_{\overline{\phi}}^{\infty}.
\]
\end{enumerate}
\end{theorem}
\Proof

\NI
$(i)$ Fix a belief model
$(\Omega, \mybar{s_1}, \LL, \mybar{s_n}, P_1, \LL, P_n)$ for $H$.
Take a strategy $s_i$ that is an element of the $i$th component of 
$G_{\textbf{RAT}({\overline{\phi}}) \cap B^* \textbf{RAT}({\overline{\phi}})}$.
Thus we have $s_i = \mybar{s_i}(\omega)$ for some state $\omega$ such that
$\omega \in \textbf{RAT}({\overline{\phi}})$ and $\omega \in B^* \textbf{RAT}({\overline{\phi}})$.
The latter implies by (\ref{equ:cb}) that for some evident event $F$
\begin{equation}
  \label{equ:F}
\omega \in F \sse \{\omega' \in \Omega \mid \fa i \in \{1, \LL, n\} \: P_i(\omega') \sse \textbf{RAT}({\overline{\phi}})\}.  
\end{equation}

Take now an arbitrary $\omega' \in F \cap
\textbf{RAT}({\overline{\phi}})$ and $i \in \{1, \LL, n\}$.  Since
$\omega' \in \textbf{RAT}({\overline{\phi}})$, player $i$ is
$\phi_i$-rational in $\omega'$, i.e., $\phi_i(\mybar{s_i}(\omega'),
G_{P_i(\omega')})$ holds.  But $F$ is evident, so $P_i(\omega') \sse
F$. Moreover by (\ref{equ:F}) $P_i(\omega') \sse
\textbf{RAT}({\overline{\phi}})$, so $P_i(\omega') \sse F \cap
\textbf{RAT}({\overline{\phi}})$.  Hence $G_{P_i(\omega')} \sse G_{F
  \cap \textbf{RAT}({\overline{\phi}})}$ and by the monotonicity of
$\phi_i$ we conclude that $\phi_i(\mybar{s_i}(\omega'), G_{F \cap
  \textbf{RAT}({\overline{\phi}})})$ holds.

By the definition of $T_{\overline{\phi}}$ this means that $G_{F \cap
  \textbf{RAT}({\overline{\phi}})} \sse
T_{\overline{\phi}}(G_{\textbf{RAT}({\overline{\phi}})})$, i.e. that
$G_{F \cap \textbf{RAT}({\overline{\phi}})}$ is a post-fixpoint of
$T_{\overline{\phi}}$.  Hence by Tarski's Fixpoint Theorem $G_{F \cap
  \textbf{RAT}({\overline{\phi}})} \sse T_{\overline{\phi}}^{\infty}$.
But $s_i = \mybar{s_i}(\omega)$ and $\omega \in F \cap
{\textbf{RAT}({\overline{\phi}})}$, so we conclude by the above
inclusion that $s_i$ is an element of the $i$th component of
$T_{\overline{\phi}}^{\infty}$.  This proves the claim.  \II

\NI
$(ii)$
By the definition of common knowledge for all events $E$ we have $K^* E \sse E$. Hence
for all $\overline{\phi}$ we have
$K^* \textbf{RAT}({\overline{\phi}}) \sse \textbf{RAT}({\overline{\phi}}) \cap K^* \textbf{RAT}({\overline{\phi}})$
and consequently
$G_{K^* \textbf{RAT}({\overline{\phi}})} \sse G_{\textbf{RAT}({\overline{\phi}}) \cap K^* \textbf{RAT}({\overline{\phi}})}$.
\II

\NI
$(iii)$
We actually construct a standard belief model  for $H$ that is also a knowledge model.
Suppose $T^{\infty}_{\overline{\phi}} = (G_1, \LL, G_n)$. Consider the event
$F := G_1 \times \LL \times G_n$ in the standard model for $H$. Then 
$G_F = T^{\infty}_{\overline{\phi}}$.
Define each possibility correspondence $P_i$ by
\[
        P_i(\omega) :=
        \left\{
        \begin{array}{l@{\extracolsep{3mm}}l}
        F    & \mathrm{if}\  \omega \in F \\
        \Omega \setminus F       & \mathrm{otherwise}
        \end{array}
        \right.
\]
Each $P_i$ is a knowledge correspondence (also when $F = \ES$ or $F = \Omega$)
and clearly $F$ is an evident event.

Take now  an arbitrary $i \in \{1, \LL, n\}$ and an arbitrary state $\omega \in F$.
Since $T^{\infty}_{\overline{\phi}}$ is a fixpoint of $T_{\overline{\phi}}$
and $\mybar{s_i}(\omega) \in G_i$ we have $\phi_{i}(\mybar{s_i}(\omega), T^{\infty}_{\overline{\phi}})$,
so by the definition of $P_{i}$ we have $\phi_{i}(\mybar{s_{i}}(\omega), G_{P_{i}(\omega)})$.
This shows that each player $i$ is $\phi_i$-rational in each state $\omega \in F$,
i.e., $F \sse \textbf{RAT}({\overline{\phi}})$. 

Since $F$ is evident, we conclude by (\ref{equ:ck}) that in each state $\omega \in F$
it is common knowledge that each player $i$ is $\phi_i$-rational, i.e.,
$F \sse K^* \textbf{RAT}({\overline{\phi}})$. Moreover, by the definition of common knowledge
$K^* \textbf{RAT}({\overline{\phi}}) \sse \textbf{RAT}({\overline{\phi}})$. 
Consequently
\[
T_{\overline{\phi}}^{\infty} = G_F \sse G_{\textbf{RAT}({\overline{\phi}}) \cap K^*
  \textbf{RAT}({\overline{\phi}})},
\]
which yields the claim by $(i)$.
\HB 
\VV

Items $(i)$ and $(ii)$ show that when each property $\phi_i$ is
monotonic, for all belief models of $H$ it holds that the strategy
profiles that the players choose in the states in which each player
$i$ is $\phi_i$-rational and it is common belief that each player $i$
is $\phi_i$-rational (or in which it is common knowledge that each
player $i$ is $\phi_i$-rational) are included in those that remain
after the iterated elimination of the strategies that are not
$\phi_i$-optimal.

Note that monotonicity of the $\phi_i$ properties was not needed to
establish item $(iii)$.  In \cite{CLL05}, \cite{Lip94} and
\cite{Apt07} examples are provided showing that for the properties of
strict dominance (namely $\textit{sd}^{\: g}$) and best response
(namely $\textit{br}^{\: g}$) in general transfinite iterations (i.e.,
iterations beyond $\omega_0$) of the corresponding operator are
necessary to reach the outcome.  So to achieve the equality for them
in $(iii)$ transfinite iterations of the $T_{\overline{\phi}}$
operator are necessary.

By instantiating $\phi_i$s to specific properties we get instances of
the above result that refer to specific definitions of rationality.
This will allow us to relate the above result to the ones established
in the literature. Before we do this we establish another result that
will apply to another class of properties $\phi_i$.

\begin{theorem} \label{thm:epist}
Suppose that
\begin{equation}
  \label{equ:1}
\mbox{$\phi_i(s_i, (\C{s_1}, \LL, \C{s_n}))$ holds for all $i \in \{1, \LL, n\}$ and all $s \in H$.}  
\end{equation}
Then for some standard knowledge model for $H$
\[
G_{K^* \textbf{RAT}(\overline{\phi})} = H.
\]
\end{theorem}

\Proof
We extend the standard model for $H$ by the knowledge correspondences $P_1, \LL, P_n$
where for all $i \in \{1, \LL, n\}$, $P_i(\omega) = \C{\omega}$.  Then for all $\omega$
and all $i \in \{1, \LL, n\}$
\[
G_{P_i(\omega)} = (\C{\mybar{s_1}(\omega)}, \LL, \C{\mybar{s_n}(\omega)}),
\]
so by (\ref{equ:1}) each player $i$ is $\phi_i$-rational in $\omega$, i.e.,
$\Omega \sse K \textbf{RAT}(\overline{\phi})$.
So, by the definition of common knowledge,
$\Omega \sse K^* \textbf{RAT}(\overline{\phi})$.
Consequently $H = G_{\Omega} \sse G_{K^* \textbf{RAT}(\overline{\phi})} \sse H$.
\HB
\VV

Note that any property $\phi_i$ that satisfies (\ref{equ:1})
and is not trivial (that is, for some strategy $s_i$, $\phi_i(s_i, H)$ does not hold)
is not monotonic.

\section{$\LL$ and their consequences}
\label{sec:consequences}

Let us analyze now the consequences of the above two theorems.
Consider first Theorem \ref{thm:epist1}.
The following lemma, in which we refer to the properties introduced
in Section \ref{sec:setup}, clarifies the matters.

\begin{lemma} \label{lem:mono}
The properties 
$sd_{i}^{\: g}, \ msd_{i}^{\: g}$ and $br_{i}^{\: g}$
are monotonic.
\end{lemma}
\Proof
Straightforward.
\HB
\VV

So Theorem \ref{thm:epist1} applies to the above three properties.
(Note that $br_{i}^{\: g}$ actually comes in three `flavours'
depending on the choice of beliefs.) Strict dominance in the
sense of $sd_{i}^{\: g}$ is studied in \cite{CLL05}, while
$br_{i}^{\: g}$ corresponds to the rationalizability notion of
\cite{Ber84}.

In contrast, Theorem \ref{thm:epist1} does not apply to the properties
$wd_{i}^{\: g}$ and $mwd_{i}^{\: g}$, since, as indicated in
\cite{Apt07c}, the corresponding operators $T_{\textit{wd}^{\: g}}$
and $T_{\textit{mwd}^{\: g}}$ are not monotonic, and hence the
properties $wd_{i}^{\: g}$ and $mwd_{i}^{\: g}$ are not
monotonic.

To see the consequences of Theorem \ref{thm:epist} note that the
properties $sd_{i}^{\: l}, \ msd_{i}^{\: l}$, $wd_{i}^{\: l}, \ 
mwd_{i}^{\: l}$ and $br_{i}^{\: l}$ satisfy (\ref{equ:1}).  In
particular, this theorem shows that the `customary' concepts of strict
dominance, $sd_{i}^{\: l}$ and $msd_{i}^{\: l}$ cannot be justified in
the used epistemic framework as `stand alone' concepts of rationality.
Indeed, this theorem shows that in some knowledge models common
knowledge that each player is rational in one of these two senses does
not exclude any strategy.

What \emph{can} be done is to justify these two concepts as
\emph{consequences} of the common knowledge of rationality defined in
terms of $br_{i}^{\: g}$, the `global' version of the best
response property, Namely, we have the following result.  When each
property $\phi_i$ equals $\textit{br}_{i}^{\: g}$, we write here 
$\textbf{RAT}({\textit{br}^{\: g}})$ instead of $\textbf{RAT}({\overline{\textit{br}^{\: g}}})$.

\begin{theorem} \label{thm:just}
\mbox{}\\[-6mm]
\begin{enumerate} \smallromani
\item For all belief models
\[
G_{\textbf{RAT}({\textit{br}^{\: g}}) \cap B^* \textbf{RAT}({\textit{br}^{\: g}})} \sse T^{\infty}_{\textit{sd}^{\: l}},
\]
\item 
for all knowledge models
\[
G_{K^* \textbf{RAT}({\textit{br}^{\: g}})} \sse T^{\infty}_{\textit{sd}^{\: l}},
\]
\end{enumerate}
where in both situations we take as the set of beliefs the set of joint strategies of the opponents.
\end{theorem}

\Proof

\NI
$(i)$
By Lemma \ref{lem:mono} and Theorem \ref{thm:epist1}$(i)$
$G_{\textbf{RAT}(\textit{br}^{\: g}) \cap B^* \textbf{RAT}(\textit{br}^{\: g})} \sse T^{\infty}_{\textit{br}^{\: g}}$.
Each best response to a joint strategy of the opponents is not
strictly dominated, so for all restrictions $G$
\[
T_{\textit{br}^{\: g}}(G) \sse T_{\textit{sd}^{\: g}}(G).
\]
Also, for all restrictions $G$
\[
T_{\textit{sd}^{\: g}}(G) \sse T_{\textit{sd}^{\: l}}(G).
\]
So by Lemma \ref{lem:inc} 
$T^{\infty}_{\textit{br}^{\: g}} \sse
T^{\infty}_{\textit{sd}^{\: l}}$, which concludes the proof.
\II

\NI
$(ii)$
By $(i)$ and the fact that 
$K^* \textbf{RAT}({\textit{br}^{\: g}}) \sse \textbf{RAT}({\textit{br}^{\: g}})$.
\HB
\VV

Item $(ii)$ formalizes and justifies in the epistemic
framework used here the often used statement:
\begin{quote}
  common knowledge of rationality implies that the players will choose
  only strategies that survive the iterated elimination of strictly
  dominated strategies
\end{quote}
for games with \emph{arbitrary strategy sets} and \emph{transfinite
  iterations} of the elimination process, and when for the set of
beliefs of a player we take the set of joint strategies of his
opponents.

In the case of finite games we have the following well-known result
implicitly stated in \cite{BD87} and explicitly formulated in
\cite{Sta94} (see \cite[ page 181]{BB99}).
For a proof using Harsanyi type spaces see \cite{BF06}.

\begin{theorem} \label{thm:just1}
Assume the initial game $H$ is finite.
\begin{enumerate} \smallromani
\item For all belief models for $H$
\[
G_{\textbf{RAT}({\textit{br}^{\: g}}) \cap B^* \textbf{RAT}({\textit{br}^{\: g}})} \sse T^{\infty}_{\textit{msd}^{\: l}},
\]
\item 
for all knowledge models for $H$
\[
G_{K^* \textbf{RAT}({\textit{br}^{\: g}})} \sse T^{\infty}_{\textit{msd}^{\: l}},
\]
\end{enumerate}
where in both situations we take as the set of beliefs the set of joint mixed strategies of the opponents.
\end{theorem}

\Proof
The argument is analogous as in the previous proof
but relies on a subsidiary result and runs as follows.

\NI
$(i)$
Again by Lemma \ref{lem:mono} and Theorem \ref{thm:epist1} 
$G_{\textbf{RAT}(\textit{br}^{\: g}) \cap B^* \textbf{RAT}(\textit{br}^{\: g})} \sse T^{\infty}_{\textit{br}^{\: g}}$.
Further, for all restrictions $G$
\[
T_{\textit{br}^{\: g}}(G) \sse T_{\textit{br}^{\: l}}(G)
\]
and
\[
T_{\textit{br}^{\: l}}(G) \sse T_{\textit{brc}^{\: l}}(G),
\]
where $brc_{i}^{\: l}$ stands for the best response property
w.r.t.~the correlated strategies of the opponents.  So by Lemma
\ref{lem:inc} $T^{\infty}_{\textit{br}^{\: g}} \sse
T^{\infty}_{\textit{brc}^{\: l}}$.

But by the result of \cite[ page 60]{OR94} (that is a modification of
the original result of \cite{Pea84}) for all restrictions $G$ we have
$T_{\textit{brc}^{\: l}}(G) = T_{\textit{msd}^{\: l}}(G)$, so
$T^{\infty}_{\textit{brc}^{\: l}} = T^{\infty}_{\textit{msd}^{\: l}}$,
which yields the conclusion.  
\II

\NI
$(ii)$ By $(i)$ and the fact that
$K^* \textbf{RAT}({\textit{br}^{\: g}}) \sse \textbf{RAT}({\textit{br}^{\: g}})$.
\HB 


\section{Axiomatic presentation}
\label{sec:axio}

It is natural to ask what proof-theoretic principles about the
players' reasoning we are assuming in the proof of Theorem
\ref{thm:epist1}$(i)$.  To answer this question we present in this section
a formal language $\cal{L}_\nu$ that will be interpreted over belief
models.  We will then give syntactic proof rules for $\cal{L}_\nu$
that lead to an axiomatic proof of Theorem
\ref{thm:epist1}$(i)$.  Throughout the section we assume, as usual, the
initial game $H$ and monotonic properties $\phi_1, \LL, \phi_n$.
Later we shall introduce a language that allows us to define and
analyze the relevant properties.

To start with, we consider the simpler language $\cal{L}$ the formulae of which are
defined by the following recursive definition, where $i \in \{1, \LL, n\}$:

\[
 \psi ::= \rat_i \mid \psi \land \psi \mid \neg \psi \mid \square_i \psi \mid \opt_i \psi, 
\]
where each $\rat_i$ is a constant.
We abbreviate the formula $\bigwedge_{i \in \{1, \LL, n\}} \rat_i$ to $\rat$, 
$\bigwedge_{i \in \{1, \LL, n\}} \square_i \psi$ to $\square \psi$ and
$\bigwedge_{i \in \{1, \LL, n\}} \opt_i \psi$ to $\opt \psi$.

Formulae of $\cal{L}$ will be interpreted as events in belief models for $H$.
Given a belief model $(\Omega, \mybar{s_1}, \LL, \mybar{s_n}, P_1, \LL, P_n)$ for $H$,
we define the \oldbfe{interpretation function} $\interp{\cdot} : {\cal L} \rightarrow {\cal P}(\Omega)$ as follows:
\begin{itemize}
 \item $\interp{rat_i} = \{ \omega \in \Omega \mid \phi_i(\mybar{s_i}(\omega), G_{P_i(\omega)})\}$,
 \item $\interp{\phi \land \psi} = \interp{\phi} \cap \interp{\psi}$,
 \item $\interp{\neg \psi} = \Omega - \interp{\psi}$,
 \item $\interp{\square_i \psi} = \{ \omega \in \Omega \mid P_i(\omega) \subseteq \interp{\psi} \}$,
 \item $\interp{\opt_i \psi} = \{ \omega \in \Omega \mid \phi_i(\mybar{s_i}(\omega), G_{\interp{\psi}})\}$.
\end{itemize}

Note that $\interp{rat}$ is the event $\textbf{RAT}(\overline{\phi})$ that every player is rational,
$\interp{\square \psi}$ is the event $\square \interp{\psi}$ that every player believes the event \interp{\psi}
and $\interp{\opt{\psi}}$ is the event that every player's strategy is optimal
 in the context of the restriction $G_{\interp{\psi}}$.
 
 $\cal{L}$ is a modal language in the sense of \cite{BRV01}.  Although
 $\cal{L}$ can express some connections between our formal definitions
 of optimality, rationality and beliefs, it is not a very expressive
 language.  If our interest were to reason about particular games, we
 could extend the language with atoms $s_i$ expressing the event that
 the strategy $s_i$ is chosen.  This choice is often made when
 defining modal languages for models of games, see, e.g.,
 \cite{dB_PHD}.  However, we are interested in a finite language that
 would allow us to reason about games with arbitrary strategy sets,
 and in particular in a language that can express the inclusion of
 Theorem \ref{thm:epist1}$(i)$.

Specifically, we want a language that can express the following statement:

\begin{itemize}
 \item[\textbf{Imp}] If it is true common belief that every player is rational,
  then all players choose strategies that survive the iterated elimination of non-optimal strategies.
\end{itemize}

To this end we extend the vocabulary of ${\cal L}$ with a single
set variable denoted by $x$ and the largest
fixpoint operator $\nu x$. 
(The corresponding extension of the first-order logic by the dual, least fixpoint operator $\mu x$ was first studied in \cite{Gur84}.)
Modulo one caveat the resulting language ${\cal L}_\nu$ is defined as follows, where `$\LL$' stands for the already 
given definition of $\cal{L}$:
\[
 \psi ::= \LL \mid x \mid \nu x. \phi.
\]

The caveat is the following: $\phi$ must be 
\begin{itemize}
\item \oldbfe{positive in} $x$, which means that 
each occurrence of $x$ in $\phi$ is under the scope of an even number of negation signs ($\neg$),

\item \oldbfe{$\nu$-free}, which means that
it does not contain any occurrences of the $\nu x$ operator.
\end{itemize}
(The latter restriction is not necessary, but simplifies matters and is sufficient for our considerations.)

To define the interpretation function $\interp{\cdot}$ for ${\cal L}_\nu$ 
we must keep track of the variable $x$.
Therefore we first extend the function $\interp{\cdot}$ on ${\cal L}$
to a function $\interp{\cdot}: {\cal L}_\nu \times {\cal P}(\Omega) \rightarrow {\cal P}(\Omega)$
by padding it with a dummy argument.
We give one clause as an example:
\begin{itemize}
 \item $\interp{\square_i \psi, E} = \{ \omega \in \Omega \mid P_i(\omega) \subseteq \interp{\psi, E} \}.$
\end{itemize}
Then we put
\begin{itemize}
 \item $ \interp{x, E} = E$,
\end{itemize}
and finally define 
   \begin{itemize}
 \item $\interp{\nu x. \psi} = \bigcup \{ E \subseteq \Omega \mid E \subseteq \interp{\psi, E} \}$.
\end{itemize}

It is straightforward to see that the restriction to
positive in $x$ and $\nu$-free formulae $\psi$ ensures that 
$\interp{\psi, \cdot}$ is a monotonic operator on the powerset ${\cal P}(\Omega)$ of $\Omega$.
Hence by Tarski's Fixpoint Theorem $\interp{\nu x. \psi}$ is its largest fixpoint.

This language can express \textbf{Imp}.
To see this, first notice that common belief is definable in ${\cal L}_\nu$
using the $\nu x$ operator.
The analogous characterization of common knowledge 
is given in \cite[ Section 11.5]{FHMV_RAK}.

\begin{note}
\label{note:expr1}
Let $\psi$ be a formula of $\cal{L}$ and $x$ a variable. Then
$\interp{\nu x.  \square (x \land \psi)}$ is the event that the event $\interp{\psi}$ is common belief.
\end{note}
\Proof
$\psi$ is a formula of $\cal{L}$, so $x$ does not occur in $\psi$. 
Note that for all $F \sse \Omega$ we have

\begin{itemize}
\item $\interp{\square (x \land \psi), F} = \interp{\square x, F} \cap \interp{\square \psi, F}$,

\item $\interp{\square x, F} = \square \interp{x, F} = \square F$, 

\item $\interp{\square \psi, F} = \square \interp{\psi, F} = \square \interp{\psi}$,
\end{itemize}
where the `outer' $\square$ is defined in 
Subsection \ref{subsec:poss}.
Hence
$\interp{\square (x \land \psi), F} = square F \cap \square \interp{\psi}$,
and consequently

\[
\begin{array}{lll}
\interp{\nu x. \square(x \land \psi)}& = & \bigcup \{ F \subseteq \Omega \mid F \subseteq \interp{\square (x \land \psi), F} \} \\ 
& = & \bigcup \{ F \subseteq \Omega \mid F \subseteq \square (F \cap \interp{\psi})\} \\ 
& = &\square^* \interp{\psi},
\end{array}
\]
where $\square^*$ is defined in Subsection \ref{subsec:poss}
and the last equality holds by Note \ref{not:alt}.
\HB
\VV

From now on we abbreviate the (well-formed) formula $\nu x. \square(x
\land \psi)$ for $\psi$ being a formula of ${\cal L}$ to $\square^*
\psi$.  So $\square^*$ is a new modality added to the language ${\cal
  L}_\nu$.

We can also define the iterated elimination of non-optimal strategies.
\begin{note} \label{note:expr2} 
In the game determined by the event $\interp{\nu x. \opt x}$, every player selects a strategy which survives the iterated elimination of non-optimal strategies.
\end{note}
\Proof We must show the following inclusion:

\[
 G_\interp{\nu x. \opt x} \subseteq T^\infty_{\overline{\phi}}.
\]

Let $G' := (G'_1, \ldots, G'_n) = G_\interp{\nu x. \opt x}$.  By Tarski's Fixpoint Theorem it
suffices to show that $G' \subseteq T_{\overline{\phi}}(G')$.
So take any $j \in \{1, \LL, n\}$ and any $s'_j \in G'_j$.  We must show that $\phi_j(s'_j, G')$ holds.
By definition for some $\omega \in \interp{\nu x. \opt x}$ we have $\mybar{s_j}(\omega) = s'_j$.  
Then there is some $E$ such that $\omega \in E$ and $E \subseteq \interp{\opt x, E}$.
Therefore for all $i \in \{1, \LL, n\}$, $\phi_i(\mybar{s_i}(\omega), G_E)$ holds, so in particular $\phi_j(s'_j, G_E)$ holds.

But $E \subseteq \interp{\opt x, E}$ implies
$E \subseteq \interp{\nu x. \opt x}$ and therefore $G_E \subseteq G_\interp{\nu x. \opt x} = G'$.  Hence
by monotonicity of $\phi_j$ we get $\phi_j(s'_j, G')$ as desired.
\HB
\VV

Now consider the following formula:
\begin{equation}
\label{eqn:syntax2}
 (\rat \land \square^* \rat) \implies \nu x. \opt x.
\end{equation}

By Notes \ref{note:expr1} and \ref{note:expr2}, we can see that wherever the formula (\ref{eqn:syntax2}) holds, then
if it is true common belief that every player is rational, then
each player selects a strategy that survives the iterated elimination of non-optimal strategies.

We call an ${\cal L}_\nu$-formula $\psi$ \oldbfe{valid} if
for every belief model $(\Omega, \mybar{s_1}, \LL, \mybar{s_n}, P_1, \LL, P_n)$ for $H$ we have
$\interp{\psi} = \Omega$.

We are now in a position to connect ${\cal L}_\nu$ to \textbf{Imp}:
the statement \textbf{Imp} asserts that the formula (\ref{eqn:syntax2}) is valid.

In the rest of this section we will discuss a simple proof system in which
we can derive (\ref{eqn:syntax2}).
This will provide an alternative way of proving the corresponding inclusion in Theorem \ref{thm:epist1}$(i)$.

We will use an axiom and rule of inference for the fixpoint operator taken from \cite{Kozen83}
and one axiom for rationality analogous to the one called in \cite{dB_PHD}
an `implicit definition' of rationality. We give these in Figure \ref{fig:proof}
denoting by $\psi[x \mapsto \chi]$
the formula obtained from $\psi$ by substituting each occurence of
the variable $x$ with the formula $\chi$.

\begin{figure}[htbp]
  \begin{center}
    \leavevmode
\fbox{
\begin{tabular}{c}
Axiom schemata \eol

\begin{tabular}{ll}

$\rat \implies (\square \psi \implies \opt \psi)$       & $\rat Dis$  \eol

$\nu x. \psi \implies \psi[x \mapsto \nu x. \psi] $     & $\nu Dis$ \eol

\end{tabular} \eol

Rule of inference \eol

$\infer[\nu Ind]{\chi \implies \nu x. \psi}{\chi \implies \psi[x \mapsto \chi]
}$ \eol

\end{tabular}
}
\caption{Proof system \textbf{P}}
\label{fig:proof}
  \end{center}

\end{figure}

First we establish the soundness of this proof system, that is that its axioms are valid
and the proof rules preserve validity.

\begin{lemma}
The proof system \textbf{P} is sound.
\end{lemma}
\Proof
We show first the validity of the axiom $\rat Dis$.
Let $(\Omega, \mybar{s_1}, \LL, \mybar{s_n}, P_i, \LL, P_n)$ be a belief model for $H$.
We must show that  $\interp{\rat \implies (\square \psi \implies \opt \psi)} = \Omega$.
That is, that for any $\psi$ the inclusion $\interp{\rat} \cap \interp{\square \psi} \subseteq \interp{O \psi}$ holds.
So take some $\omega \in \interp{\rat} \cap \interp{\square \psi}$.
Then for every $i \in \{1, \LL, n\}$, $\phi_i(\mybar{s_i}(\omega), G_{P_i(\omega)})$,
and $P_i(\omega) \subseteq \interp{\psi}$.
So by monotonicity of $\phi_i$, $\phi_i(\mybar{s_i}(\omega), G_{\interp{\psi}})$, i.e.~$\omega \in \interp{\opt_i \psi}$ as required.

The axioms $\nu Dis$ and the rule $\nu Ind$ were introduced in \cite{Kozen83}, and their soundness proof is standard. 
This axiom and the rule formalize, respectively, the following two consequences of Tarski's Fixpoint Theorem
concerning a monotonic operator $T$:
\begin{itemize}
\item $\nu T$ is a post-fixpoint of $T$, i.e., $\nu T \sse T(\nu T)$ holds,

\item if $Y$ is a post-fixpoint of $T$, i.e., $Y \sse T(Y)$ holds, then
$Y \sse \nu T$.
\HB
\end{itemize}
\VV

Next, we establish the already announced claim.
\begin{theorem}
The formula (\ref{eqn:syntax2}) is a theorem of the proof system \textbf{P}.
\end{theorem}
\Proof
The following formula is an instance of the axiom $\rat Dis$ (with $\psi := \square^* \rat \land \rat$):

\[
 \rat \implies (\square (\square^* \rat \land \rat) \implies \opt(\square^* \rat \land \rat)),
\]
and the following is an instance of $\nu Dis$ (with $\psi := \square (x \land \rat)$):

\[
 \square^* \rat \implies \square(\square^* \rat \land \rat)
\]

Putting these two together via some simple propositional logic, we obtain:

\[
 (\square^* \rat \land \rat) \implies \opt (\square^* \rat \land \rat).
\]

This last formula is of the right shape to apply the rule $\nu Ind$
(with $\chi := \square^* \rat \land \rat$ and $\psi := \opt x$), to obtain:

\[
 (\square^* \rat \land \rat) \implies \nu x. \opt x,
\]
which is precisely the formula (\ref{eqn:syntax2}).
%
\HB
\VV

The derivation of (\ref{eqn:syntax2}) has shown which proof-theoretic
principles are sufficient to obtain Theorem \ref{thm:epist1}$(i)$.  It is
interesting to note that no axioms or rules for the modalities
$\square$ and $\opt$ were needed in order to derive
(\ref{eqn:syntax2}).  This corresponds to the fact that in the proof
of the corresponding inclusion in Theorem \ref{thm:epist1}$(i)$ we did not
use the fact that the possibility correspondences were belief
correspondences.

\begin{corollary}
The formula (\ref{eqn:syntax2}) is valid. 
\HB
\end{corollary}

In the language ${\cal L}_\nu$, $\rat_1, \LL, \rat_n$ are
propositional constants.  We can define them in terms of the
$\square_i$ and $\opt_i$ modalities but to this end we need to extend
the language ${\cal L}_\nu$ to a second-order one by allowing
quantifiers over set variables, so by allowing formulae of the form
$\exists X \phi$. It is clear how to extend the semantics to this
larger class of formulae.  In the resulting language each $\rat_i$
constant is definable by a formula of the latter language:
\begin{equation} 
  \label{equ:2nd}
\rat_i \equiv \forall X (\square_i X \implies \opt_i X),   
\end{equation}
where $\forall X \phi$ is an abbreviation for $\neg \exists X \neg \phi$.

The following observation then shows correctness of this definition.

\begin{note} For all $i \in \{1, \LL, n\}$
the formula (\ref{equ:2nd}) is valid. 
\HB
\end{note}

Let us mention that such second-order extensions of propositional
modal logics were first considered in \cite{Fin70}.

To further our syntactic analysis, we now give a language $\mathcal{L}_O$
which can be used to define and analyze the optimality properties $\phi_i(\cdot,
\cdot)$.  It is a first-order language formed from a family of $n$ ternary relation
symbols $x \geq_z^i y$, where $i \in \{1, \LL, n\}$,
along with the binary relation $x \in X$ between a first-order variable and a set variable.
$\mathcal{L}_O$ is given by the following recursive definition:

\[
 \phi ::= x \in X \mid x \geq_z^i y \mid \neg \phi \mid \phi \A \phi \mid \exists x \phi,
\]
where $i \in \{1, \LL, n\}$.

We use the same abbreviations $\implies$ and $\lor$ as above and further
abbreviate $\neg \: y \geq_z^i x$ to $x >_z^i y$, $\te x (x \in X \A
\phi)$ to $\te x \in X \: \phi$, $\fa x (x \in X \myra \phi)$ to $\fa
x \in X \: \phi$, and write $\forall x \phi$ for $\neg \exists x \neg
\phi$.

By an \oldbfe{optimality condition} for player $i$ we now mean a
formula containing exactly one free first-order variable and the set variable
$X$, and in which all the occurrences of the atomic formula $x \geq_z^j y$
are with $j$ equal to $i$.  In
particular, we are interested in the following optimality conditions:

\begin{itemize}
 \item $sd^l_{i}(x, X) := \forall y \in X \exists z \in X x \geq_z^i y$,

 \item $sd^g_{i}(x, X) := \forall y \exists z \in X x \geq_z^i y$,

 \item $wd^l_{i}(x, X) := \forall y \in X (\forall z \in X x \geq_z^i y \lor \exists z \in X x >_z^i y)$,

 \item $wd^g_{i}(x, X) := \forall y ( \forall z \in X x \geq_z^i y \lor \exists z \in X x >_z^i y)$,

 \item $br^l_{i}(x, X) := \exists z \in X \forall y \in X x \geq_z^i y$,

 \item $br^g_{i}(x, X) := \exists z \in X \forall y \: x \geq_z^i y$.
\end{itemize}

We now give a semantics for $\mathcal{L}_O$-formulae in the context of
a model $(\Omega, \mybar{s_1}, \ldots, \mybar{s_n})$ for the initial game $H$. 
An \oldbfe{assignment} is a function $\alpha$ that maps each first-order
variable to a state in $\Omega$ and each set variable to an event in (a subset
of) $\Omega$. The semantics is given by a satisfaction relation
between an assignment $\alpha$ and a formula $\phi$ of 
$\mathcal{L}_O$, with $\models_{\alpha} \phi$ meaning that $\alpha$ satisfies $\phi$.
This relation is defined as follows:

\begin{itemize}
 \item $\models_\alpha x \in X$ iff $\alpha(x) \in \alpha(X)$,
 \item $\models_\alpha x \geq_z^i y$ iff $p_i(\mybar{s_i}(\alpha(x)), \mybar{s_{-i}}(\alpha(z))) \geq p_i(\mybar{s_i}(\alpha(y)), \mybar{s_{-i}}(\alpha(z)))$,
 \item $\models_\alpha \neg \phi$ iff not $\models_\alpha \phi$,
 \item $\models_\alpha \phi \A \psi$ iff $\models_\alpha \phi\textup{ and } \models_\alpha \psi$,
 \item $\models_\alpha \exists x \phi$ iff $\textup{there is an }\omega \in \Omega\textup{ such that } \models_{\alpha[x \mapsto \omega]} \phi$,
\end{itemize}
where:
\[
\alpha[x \mapsto \omega](x_0) := \left\{
\begin{array}{ll}
 \alpha(x) & \textup{ if } x \neq x_0 \\
 \omega & \textup{ otherwise}.
\end{array}
\right.
\]

This semantics allows us to relate the above six optimality conditions
to the corresponding optimality properties that are
concerned solely with pure strategies.

\begin{note}
For each optimality condition $\phi_i$, where $\phi \in \{sd^l, \: sd^g, wd^l, \: wd^g, br^l, \: br^g\}$
\[
\mbox{$\models_\alpha \phi_i(x, X)$ iff the property $\phi_i(\mybar{s_i}(\alpha(x)), G_{\alpha(X)})$ holds.} 
\]
\HB
\end{note}

To relate optimality conditions to monotonic optimality properties we need
one more definition.  We say that a formula $\phi$ of $\mathcal{L}_O$ 
is \emph{positive}
just when every occurrence of the set variable $X$ occurs under a
positive number of negation signs ($\neg$).  So for example the
formula $br^l_{i}(x, X)$, that is, $\exists z \in X \forall y \in X x
\geq_z^i y$, is not positive, since the second occurrence of $X$ is
under one negation sign, while $br^g_{i}(x, X)$, that is, $\exists z
\in X \forall y \: x \geq_z^i y$, is positive.

The following observation then links syntactic matters with 
monotonicity.

\begin{note}
For every positive optimality condition $\phi_i(x, X)$ for player $i$
the corresponding property $\phi_i(s_i, G)$ (used by player $i$) is monotonic.
\HB
\end{note}

Among the above six optimality conditions only $sd^g_{i}(x, X)$ and
$br^g_{i}(x, X)$ are positive. The corresponding other four
properties, as already mentioned earlier, are not monotonic.  By the
above observation they cannot be defined by positive formulae.

\section{Public announcements}
\label{sec:public}

The main result, Theorem \ref{thm:epist1}$(i)$, dealt with the outcome
$T_{\overline{\phi}}^{\infty}$ of the iterated elimination of
strategies that for player $i$ are not $\phi_i$-optimal, and crucially
relied on the assumption that each property $\phi_i$ is monotonic.
However, this outcome exists for arbitrary $\phi_i$s.  In this section
we show that for a large class of properties $\phi_i$ this outcome can
be characterized by means of the concept of a \emph{public
  announcement}. This approach, inspired by \cite{vB07}, applies to
all global properties introduced in Section \ref{sec:setup}, some of
which are non-monotonic.

The particular kind of ``public announcement'' that we will be
interested in is a set of true statements, one by each player $i$ to
the effect that $i$ will not play any strategy that is not optimal for
him, according to his notion of optimality.  Note that there is no
strategic element to these announcements: the players simply follow a
protocol from which they cannot deviate.  The announcements are
``public'' in the sense that every other player hears them as they
happen.

The iterated public announcements can be thought of as a process in
which the players \emph{learn} how the game will be played.  The limit
of this learning process represents the situation in which the
announcements lead to no change in the model, at which point it can be
said that rationality has been learned by all players. It is in this
sense that public announcements provide alternative epistemic
foundations for the outcome $T_{\overline{\phi}}^{\infty}$.

Let us clarify first what we would like to achieve.  Consider a model
${\cal M}$ for the initial game $H$. The process of iterated
elimination of the strategies that are not $\phi_i$-optimal,
formalized by the iterated applications of the $T_{\overline{\phi}}$
operator, produces a sequence $T^{\alpha}_{\overline{\phi}}$, where
${\alpha}$ is an ordinal, of restrictions of $H$. We would like to
mimic it on the side of the models, so that we get a corresponding
sequence ${\cal M}^{\alpha}$ of models of these restrictions.

To make this idea work we need to define an appropriate way of
reducing models. We take care of it by letting the players repeatedly
announce that they only select $\phi_i$-optimal strategies.  This
brings us to the notions of public announcements and their effects on
the models.

Given a model ${\cal M} = (\Omega, \mybar{s_1}, \LL, \mybar{s_n})$
we define 

\begin{itemize}

\item a \oldbfe{public announcement} by player $i$ in a model ${\cal M}$
as an event $E$ in ${\cal M}$,

\item given a vector $\overline{E} := (E_1, \LL, E_n)$ of public announcements by
players $1, \LL, n$
we let
\[
[\overline{E}]({\cal M}) := (\cap_{i = 1}^{n} E_i, (\mybar{s_i} \restr{\cap_{i = 1}^n
E_i})_{i \in \{1, \LL, n\}})
\]
and call it \oldbfe{the effect of the public announcements of
  $\overline{E}$ on ${\cal M}$}.
\end{itemize}

Given a property $\phi_i(\cdot, G)$ that player $i$ uses to select his
strategies in the restriction $G$ of $H$ and a model ${\cal M} :=
(\Omega, \mybar{s_1}, \LL, \mybar{s_n})$ for $G$ we define $\newMS{\phi_i}$ as the
event in ${\cal M}$ that player $i$ selects optimally his strategies
with respect to $G$.  Formally:
\[
\newMS{\phi_i} := \{\omega \in \Omega \mid \phi_i(\mybar{s_{i}}(\omega), G)\}
\]
(Note that in the notation of the previous section we have
$\newMS{\overline{\phi}} = \interp{\opt_i \textbf{T}}$,
where $\textbf{T} := \psi \lor \neg \psi$.)  We abbreviate the vector
$(\newMS{\phi_1}, \LL, \newMS{\phi_n})$ to $\newMS{\overline{\phi}}$.

We want now to obtain the reduction of a model ${\cal M}$ of $G$ to a
model ${\cal M}$ of $T_{\overline{\phi}}(G)$ by means of the just
defined vector $\newMS{\overline{\phi}}$ of public announcements.  

The effect of the public announcements of $\overline{E}$ on a model of
$G$ should ideally be a model of the restriction $G_{\overline{E}}$.
Unfortunately, this does not hold in such generality.  Indeed, let the
two-player game $G$ have the strategy sets $G_1 := \{U, D\}$, $G_2 :=
\{L, R\}$ and consider the model $\mathcal{M}$ for $G$ with $\Omega :=
\{\omega_{ul}, \omega_{dr}\}$ and the functions $\mybar{s_1}$ and $\mybar{s_2}$
defined by
\[
\mybar{s_1}(\omega_{ul}) = U, \ \mybar{s_2}(\omega_{ul}) = L, \ \mybar{s_1}(\omega_{dr}) = D, \ \mybar{s_2}(\omega_{dr}) = R.
\]
This simple example is depicted in Figure \ref{fig:ceg}.

\begin{figure}[htbc]
\begin{center}
\begin{game}{2}{2}
        & $L$                & $R$                \\
$U$        & $\omega_{ul}$        &                 \\
$D$        &                 & $\omega_{dr}$
\end{game}
\caption{\label{fig:ceg}A motivating example for the use of standard models}
\end{center}
\end{figure}

Let $\overline{E} = (\{\omega_{ul}\}, \{\omega_{dr}\})$; then
$[\overline{E}](\cal{M}) = \emptyset$, which is not a model of $G_{\overline{E}} =
(\{U\}, \{R\})$.

A remedy lies in restricting one's attention to the standard models.
However, in order to find a faithful public announcement analogue to strategy elimination
we must also
narrow the concept of a public announcement as follows.
A \oldbfe{proper public announcement} by player $i$ in a
standard model is a subset of
$\Omega = G_1 \times \LL \times G_n$ of the form $G_1 \times \LL
\times G_{i-1} \times G'_i \times G_{i+1} \times \LL \times G_n$.

So a proper public announcement by a player is an event that amounts
to a `declaration' by the player that he will limit his attention to a
subset of his strategies, that is, will discard the remaining
strategies.  So when each player makes a proper public announcement,
their combined effect on the standard model is that the set of states
(or equivalently, the set of joint strategies) becomes appropriately
restricted.  An example, which is crucial for us, of a proper public
announcement in a standard model is of course $\newMS{\phi_i}$.

The following note links in the desired way two notions we introduced.
It states that the effect of the proper public announcements of
$\overline{E}$ on the standard model for $G$ is the standard model
for the restriction of $G$ to $\overline{E}$.

\begin{note} \label{not:link2}
  Let ${\cal M}$ be the standard model for $G$ and $\overline{E}$ a
  vector of proper public announcements by players $1, \LL, n$ in
  ${\cal M}$.  Then $[\overline{E}]({\cal M})$ is the standard model
  for $G_{\overline{E}}$.
\end{note}

\Proof
We only need to check that $\cap_{i = 1}^{n} E_i$ is the set of joint strategies of
the restriction $G_{\overline{E}}$.
But each $E_i$ is a proper announcement, so it is of the form 
$G_1 \times \LL \times G_{i-1} \times G'_i \times G_{i+1} \times \LL \times G_n$, 
where $G = (G_1, \LL, G_n)$. So $\cap_{i = 1}^{n} E_i = G'_1 \times \LL \times G'_n$. 

Moreover, each function $\mybar{s_i}$ is a projection, so
$G_{\overline{E}} = (\mybar{s_1}(E_1), \LL, \mybar{s_n}(E_n)) = (G'_1, \LL, G'_n)$. 
\HB
\VV

We also have the following observation that links the vector
$\newMS{\overline{\phi}}$ of public announcements with the operator
$T_{\overline{\phi}}$ of Section \ref{sec:setup}.

\begin{note} \label{not:link}
  Let ${\cal M} := (\Omega, \mybar{s_1}, \LL, \mybar{s_n})$ be the standard model for $G$. Then
\[
T_{\overline{\phi}}(G) = G_{[\![\overline{\phi}]\!]}.
\]
\end{note}
\Proof
Let $G = (G_1, \LL, G_n)$, $T_{\overline{\phi}}(G) = (G'_1, \LL, G'_n)$ and
$G_{[\![\overline{\phi}]\!]} = (S''_1, \LL, S''_n)$.

Fix $i \in \{1, \LL, n\}$. Then we have the following string of equivalences:
\[
\begin{array}{lll}
&              & s_i \in G'_i  \\
& \textrm{iff} & s_i \in G_i \A \phi_i(s_i, G) \\
\textrm{($\mybar{s_i}$ is onto)} & \textrm{iff} & s_i \in G_i \A \te \omega \in \Omega \: (s_i = \mybar{s_i}(\omega) \A \phi_i(\mybar{s_i}(\omega), G)) \\
& \textrm{iff} & s_i \in G_i \A \te \omega \in \newMS{\phi_i} \: (s_i = \mybar{s_i}(\omega)) \\
& \textrm{iff} & s_i \in S''_i.
\end{array}
\]
\HB
\VV

Denote now by $\newMS{\overline{\phi}}^{\infty}$ 
the iterated effect of the public
announcements of $\newMS{\overline{\phi}}$ 
starting with the standard model
for the initial game $H$.
The following conclusion then relates the iterated elimination of the
strategies that for player $i$ are not $\phi_i$-optimal to the iterated effects of
the corresponding public announcements.

\begin{corollary}
\label{cor:optanc}
$\newMS{\overline{\phi}}^{\infty}$ is the standard model for the
restriction $T_{\overline{\phi}}^{\infty}$.
\end{corollary}
\Proof
By Notes \ref{not:link2} and \ref{not:link}.
\HB
\VV

Note that in the above corollary each effect of the
public announcements of $\newMS{\overline{\phi}}$ is considered on
a different standard model.  Note also that the above result
holds for arbitrary properties $\phi_i$, not necessarily monotonic
ones.

We already mentioned in Section \ref{sec:intro} that for various
natural properties $\overline{\phi}$ transfinite iterations of
$T_{\overline{\phi}}$ may be needed to reach the outcome
$T_{\overline{\phi}}^{\infty}$. So the same holds for the iterated effects of
the corresponding public announcements.  It is useful to point out
that, as shown in \cite{Par92} a similar situation can arise in case of
natural dialogues the aim of which is to reach common knowledge.

This analysis gives an account of public announcements of the
\emph{optimality} of players' strategies.  We now extend this analysis to
public announcements of \emph{rationality}.  To this end we
additionally assume for each player a belief correspondence $P_i:
\Omega \rightarrow {\cal P}(\Omega)$, that is we consider belief
models.

We define then the event of player $i$ being $\phi_i$-rational in the
restriction $G$ as
\[
\eventra{\phi_i} := \{\omega \in \Omega \mid \phi_i(\mybar{s_{i}}(\omega), G_{P_i(\omega)})\}.
\]
(Note that in the notation of the previous section we have
$\eventra{\phi_i} = \interp{\rat_i}$.)
Again we abbreviate $(\eventra{\phi_1}, \ldots, \eventra{\phi_n})$ to
$\eventra{\overline{\phi}}$.
Note that $\eventra{\overline{\phi}}$ depends on the underlying belief model
$(\Omega, \mybar{s_1}, \LL, s_n, P_1, \LL, P_n)$ and on $G$. 

We extend the definition of the effect of the public announcements
$\overline{E} := (E_1, \LL, E_n)$ to belief models in the natural way,
by restricting each possibility correspondence to the intersection of
the events in $\overline{E}$:
\[
 [\overline{E}](\mathcal{M}, P_1, \LL, P_n) = ([\overline{E}]\mathcal{M}, P_1
\restr{\cap_{i = 1}^nE_i}, \LL, P_n \restr{\cap_{i = 1}^nE_i}).
\]
This definition is in the same spirit as in \cite{Pla89}
and in \cite[ page 72]{OR94}, where it is used
in the analysis of the puzzle of the hats.

We aim to find a class of belief models for which, under a mild
restriction on the properties $\phi_i$, $\langle
\overline{\phi}\rangle^\infty$, the iterated effect of the public
announcements of $\langle \overline{\phi}\rangle$ starting with the
standard belief model for the initial game $H$, will be the standard
belief model for $T^\infty_{\overline{\phi}}$.  We will therefore use
a natural choice of possibility correspondences, which we call the
\oldbfe{standard possibility correspondences}:
\[
 P_i(\omega) = \{\omega' \in \Omega \mid \mybar{s_i}(\omega) = \mybar{s_i}(\omega')\}.
\]
By the \oldbfe{standard knowledge model} for a restriction  
$G$ we now mean the standard model for $G$ endowed with
the standard possibility correspondences.

The following observation holds.

\begin{note} \label{not:poss}
Consider the standard knowledge model $(\Omega, \mybar{s_1}, \LL, \mybar{s_n}, P_1, \LL, P_n)$
for a restriction $G := (G_1, \LL, G_n)$ of $H$ and a state
$\omega \in \Omega$. Then
\[
G_{P_i(\omega)} = (G_1, \LL, G_{i-1}, \{\omega_i\}, G_{i+1}, \LL,  G_n).
\]
\end{note}
\Proof
Immediate by the fact that in the standard knowledge model for each
possibility correspondence we have
$P_i(\omega) = \{\omega' \in \Omega \mid \omega'_i = \omega_i\}$.
\HB
\VV

Intuitively, this observation states that in each state of a standard
knowledge model each player knows his own choice of strategy but knows
nothing about the strategies of the other players.  So standard
possibility correspondences represent the beliefs of each player after
he has privately selected his strategy but no information between the
players has been exchanged.  It is in that sense that the standard
knowledge models are natural.  In \cite{vB07} in effect only such
models are considered.

A large class of properties $\phi_i$ satisfy the following restriction:
\begin{description}
 \item[A] For all $G := (G_1, \ldots, G_n)$ and $G' := (G'_1, \ldots, G'_n)$
such that $G_j = G_j'$ for all $j \neq i$,

$\phi_i(s_i, G)\: \lra \: \phi_i(s_i, G')$.
\end{description}

That restriction on the properties 
$\phi_i$ is sufficient to obtain the following analogue of
Corollary \ref{cor:optanc} for the case of public 
announcements of \emph{rationality}.

\begin{theorem}
\label{thm:ratanc}
Suppose that each property $\phi_1, \LL, \phi_n$ satisfies \textbf{A}.
Then $\eventra{\overline{\phi}}^\infty$ is the standard knowledge
model for the restriction $T^\infty_{\overline{\phi}}$.
\end{theorem}
\Proof
Notice that it suffices to prove for each restriction $G$
the following statement for each $i$:

\begin{equation}
\forall \omega \in \Omega \:\phi_i(\omega_i, G)\:\lra\:\phi_i(\omega_i,
G_{P_i(\omega)}).
  \label{equ:equal}
\end{equation}

Indeed, (\ref{equ:equal}) entails that $\eventra{\phi_i} =
\newMS{\phi_i}$, in which case the result follows from Corollary
\ref{cor:optanc} and the observation that the possibility
correspondences are restricted in the appropriate way.

But (\ref{equ:equal}) is a direct consequence of the assumption of \textbf{A}
and of Note \ref{not:poss}.
\HB
\VV

To see the consequences of the above result
note that \textbf{A} holds for each global property
$\textit{sd}_{i}^{\: g}, \ \textit{msd}_{i}^{\: g}$, $\textit{wd}_{i}^{\: g}, \ 
\textit{mwd}_{i}^{\: g}$ and (all three forms of)
$\textit{br}_{i}^{\: g}$ introduced in Section
\ref{sec:setup}.  

For each $\phi_i$ equal $\textit{sd}_{i}^{\: g}$ and finite games
Theorem \ref{thm:ratanc} boils down to Theorem 7 in \cite{vB07}. The
corresponding result for each $\phi_i$ equal to $\textit{br}_{i}^{\: g}$,
with the beliefs consisting of the joint strategies of the opponents,
and finite games is mentioned at the end of Section 5.4 of that paper.

It is interesting to recall that the properties 
$\textit{wd}_{i}^{\: g}$ and $\textit{mwd}_{i}^{\: g}$, in contrast to
$\textit{sd}_{i}^{\: g}$ and $\textit{msd}_{i}^{\: g}$ and
$\textit{br}_{i}^{\: g}$, are not monotonic.
So, in contrast to Theorem \ref{thm:epist1}, we have now a characterization of
$T^\infty_{\overline{\phi}}$ for both forms of weak dominance.

Also it is important to note that the above 
Corollary does not hold for the
corresponding local properties $\textit{sd}_{i}^{\: l}, \ \textit{msd}_{i}^{\:
  l}$, $\textit{wd}_{i}^{\: l}, \ \textit{mwd}_{i}^{\: l}$ and
$\textit{br}_{i}^{\: l}$ introduced in Section \ref{sec:setup}.
Indeed, for each such property $\phi_{i}$ 
by Note \ref{not:poss} 
$\phi_{i}(\omega_i, G_{P_i(\omega)})$ holds for each state $\omega$
and restriction $G$. 
Consequently $\eventra{\overline{\phi}} = \Omega$. 
So when each $\phi_i$ is a local property listed above,
$\eventra{\overline{\phi}}$ is an identity operator on the standard
knowledge models, that is
$\eventra{\overline{\phi}}^\infty$ is the standard knowledge
model for the initial game $H$ and not $T^\infty_{\overline{\phi}}$.

Still, as the following result shows, it is possible for finite games to draw
conclusions about the outcome of the iterated elimination of strategies
that are not optimal in a local sense.

\begin{theorem}
\label{cor:ratanc1}
Assume the initial game $H$ is finite.  Then for each $\phi \in \{sd,
msd, wd, mwd, br\}$, $\eventra{\phi^{g}}^\infty$ is the standard
knowledge model for the restriction $T^\infty_{\phi^{l}}$.
\end{theorem}
\Proof
We rely on the following results that for finite games link the outcomes of the iterations of 
the corresponding local and global properties:
 \begin{itemize}
 \item (see \cite{Apt07})

 $T^{\infty}_{{br}^{\: l}} = T^{\infty}_{{br}^{\: g}}$,

 \item (see \cite{Apt07c})

$T^{\infty}_{{\phi}^{l}} = T^{\infty}_{{\phi}^{g}}$
 for $\phi \in \{sd, wd\}$,

 \item (see \cite{BFK06})

$T^{\infty}_{{msd}^{l}} = T^{\infty}_{{msd}^{g}}$,

 \item (see \cite{BFK08})

$T^{\infty}_{{mwd}^{l}} = T^{\infty}_{{mwd}^{g}}$.
 \end{itemize}

The conclusion now follows by Theorem \ref{thm:ratanc}.
\HB
\VV

This corollary states that for finite games the outcome of, for
example the customary iterated elimination of weakly dominated
strategies, $T^{\infty}_{{wd}^{l}}$, can be obtained by iterating on
the standard knowledge models the effect of the public announcements
by all players of the corresponding \emph{global version} of weak dominance,
${wd}_{i}^{\: g}$.
So, yet again, we see an intimate interplay between the local and global
notions of dominance.

\section*{Acknowledgements}

We acknowledge helpful discussions with Adam Brandenburger, who
suggested Theorems \ref{thm:just} and \ref{thm:just1}, and with
Giacomo Bonanno who, together with a referee of \cite{Apt07}, suggested
to incorporate common beliefs in the analysis.  Joe Halpern pointed us
to \cite{MS89}.  Johan van Benthem made us aware of the alternative
approach to epistemic analysis based on public announcements.
Finally, we thank the referee for most helpful and extensive comments.

\bibliography{/ufs/apt/bib/e,/ufs/apt/bib/apt}
\bibliographystyle{handbk}

\end{document}